\documentclass[twocolumn,trackchanges]{aastex701}

\usepackage{amsmath}
\usepackage{xcolor}

\hypersetup{linkcolor=red,citecolor=blue,filecolor=cyan}

\newcommand{\msun}{\mathrm {M_\odot}}
\newcommand{\mbh}{M_{\rm BH}}
\newcommand{\ledd}{\lambda_{\rm Edd}}
\newcommand{\rschw}{R_{\rm Schw}}
\newcommand{\ab}{\alpha_{\rm B}}
\usepackage{soul}

\usepackage[T1]{fontenc}
\usepackage{float}

\begin{document}

\title{A magnetically-supported disk-corona model for Changing-Look AGN transitions}

\author[orcid=0009-0005-4351-2801,sname='Kouzis']{Marios Kouzis}
\affiliation{Nicolaus Copernicus Astronomical Center, Polish Academy of Sciences, Bartycka 18, 00716 Warsaw, Poland}
\email[show]{makouz@camk.edu.pl}  

\author[orcid=0000-0002-5275-4096, sname='Różańska']{Agata Różańska} 
\affiliation{Nicolaus Copernicus Astronomical Center, Polish Academy of Sciences, Bartycka 18, 00716 Warsaw, Poland}
\email{agata@camk.edu.pl}

\author[orcid=0000-0003-0826-9787, sname=Lančová]{Debora Lančová}
\affiliation{Nicolaus Copernicus Astronomical Center, Polish Academy of Sciences, Bartycka 18, 00716 Warsaw, Poland}
\affiliation{Research Center for Computational Physics and Data Processing, Institute of Physics, Silesian University in Opava, Bezručovo nám. 13, 746 01 Opava, Czech Republic}
\email{debora.lancova@physics.slu.cz}

\author[orcid=0000-0001-5848-4333, sname='Czerny']{Bożena Czerny} 
\affiliation{Center for Theoretical Physics, Polish Academy of Sciences, Al. Lotników 32/46, 02-668 Warsaw, Poland}
\email{bcz@cft.edu.pl}

\author[orcid=0000-0001-5325-7705, sname='Gronkiewicz']{Dominik Gronkiewicz} 
\affiliation{Space Research Centre, Polish Academy of Sciences, Solar Physics Division, Kopernika 11, 51-622 Wrocław, Poland}
\email{dg@cbk.pan.wroc.pl}

\begin{abstract}
Changing-Look Active Galactic Nuclei (CLAGN) undergo dramatic spectral and luminosity transitions on timescales of months to a few years -- orders of magnitude shorter than the viscous timescale of a standard $\alpha$-disk at the radii where the optical/UV continuum is generated, for typical supermassive black hole masses. We show that a magnetically supported disk--corona model reproduces \emph{both} the observed Eddington ratio at which changing event occurs and the observed transition duration. Using the \texttt{diskvert} code, which solves the steady vertical structure under simultaneous gas, radiation and magnetic pressure support with a self-consistent warm corona, we (i) construct thermal-viscous S-curves, and (ii) calculate the integrated thermal timescale  together with the front propagation timescale. We compute a large grid of models of different black hole masses, Eddington ratios, magnetic viscosities, and disk radii, showing that magnetized disks push the S-curve knee down to an Eddington ratio of $ \approx 0.01\text{--}0.03$, and introduce a new stable branch of high luminosity solutions, while the limit-cycle timescale enters the months-to-years range for $\mbh = 10^{7}\text{--}10^{9}\,\msun$. Confronted with a sample of five CLAGN (Mkn 590, NGC 1566, IRAS 23226$-$3843, Mkn 1018, NGC 2617), the model jointly reproduces the empirical Eddington rates and the observed event durations only when the inner disk is strongly magnetized. The case of Mkn 590 is especially constraining: the recent tightly-determined transition Eddington ratio is matched by a highly magnetised disk-corona flow at small radii. 

\end{abstract}

\keywords{\uat{Active galactic nuclei}{16} --- \uat{Accretion}{14} --- \uat{Supermassive black holes}{1663} --- \uat{High energy astrophysics}{739}}

\section{Introduction} \label{sec:intro} 
Changing-Look Active Galactic Nuclei (CLAGN) are sources whose optical, ultraviolet and X-ray continuum, together with the broad emission-line complex, change qualitatively on timescales as short as months \citep{Trakhtenbrot19, Zeltyn22} and as long as a few years \citep[see review by][and references therein]{Ricci23}. The spectral type, defined on the bases of the AGN unified model \citep{Antonucci93,Urry1995PASP..107..803U}, can change between Type-1 and Type-2, while the observed bolometric luminosity can vary by more than an order of magnitude. Several sources have now been observed to repeat the cycle \citep[e.g.][]{Wang24}. The associated Eddington ratio, $\ledd \equiv L_\mathrm{bol} / L_\mathrm{Edd}$, at which the transition takes place is empirically constrained to be low \citep[e.g.][]{Noda18,panda2024}. A recent uniform analysis of a CLAGN sample places the critical value near $\log\lambda_{\rm Edd} \approx -2$ \citep{Jana25,Jana26}, and the dedicated long-baseline study of Mkn 590 by \citet{Palit26} narrows the transition to $\log\lambda_{\rm Edd} = -1.68^{+0.14}_{-0.21}$.

Interestingly, the observed duration of the CL transition, from month to a few years, cannot be explained by the classical $\alpha$-disks \citep{SS73}. This is because the timescale on which the surface density evolves is the viscous one, $t_{\rm vis}\sim r^{2}/\nu$. At the radii where the optical/UV continuum is generated ($\sim 10^3 \ \mathrm{R_g}$), it reaches $\sim 500$~yr for a $10^{8}\, \msun$ black hole \citep{Lawrence18}  - two to three orders of magnitude longer than what is observed. 
Several explanations have been proposed: a limit-cycle radiation-pressure instability operating in a narrow ring at the boundary between an inner advection-dominated flow and an outer standard disk, with the cycle timescale set by the radial narrowness of the unstable zone and further reduced by magnetic outflows \citep{Sniegowska20,Pan21}; cooling/heating-front propagation in the inner accretion flow \citep{Hameury09}; rapid magnetic flux inversion in a magnetically arrested disk \citep{Scepi21}; off-axis tidal disruption events or stars embedded in the disk \citep{Merloni15,Trakhtenbrot19}; and warps in the disk \citep{Raj21}. Each can in principle reproduce a subset of the observed phenomenology, but no model has so far simultaneously replicated the empirical CL transition Eddington ratios 
and the observed durations across a sample, from a single self-consistent set of disk solutions.

Magnetically supported disks offer a natural framework in which both diagnostics can be addressed. Simulations and analytical modelling have repeatedly shown that  the magnetic field, which drives the angular momentum transport via magneto-rotational instability (MRI), can generate magnetic pressure which dominates the disk pressure budget over the gas and radiation pressures \citep{Begelman07, Dexter19, Lancova19, Begelman24}. Further, these strongly magnetized disks are stable against the thermal and viscous instabilities, which are known to arise in the \cite{SS73} thin disk model when the radiation pressure starts to dominate \citep{Lightman74, SS76}. Independently, the soft X-ray excess detected in most Type-1 AGN is well described by emission from a warm ($kT \sim 0.1$-$1$ keV), optically thick ($\tau \sim 10$-$30$) layer covering the cold disk \citep{Petrucci18,Petrucci20}. Such a warm corona arises self-consistently in vertically resolved solution of magnetically supported disk, which was introduced in \citet{Gronkiewicz20} and \citet{Gronkiewicz23}, and which we also adopt here.

The stabilizing role of magnetic pressure is most cleanly seen on the thermal-equilibrium curve in the $\lambda_{\rm Edd}$ and column density, $\Sigma$, plane: an S-curve traces the location of thermally balanced solutions, and the knee separates the stable upper branch from the stable lower branch through a thermally unstable middle segment. The Eddington ratio at the knee is the local trigger threshold for a state transition, and its location depends on which pressure component dominates the vertical structure \citep{SS76,Svensson94}.

The standard Shakura-Sunyaev disks solution has no stable radiation-pressure dominated upper branch, unless the advection cooling is accounted for, as in the slim-disk model of \citet{Abramowicz88}. However, our model does not include advection, and the stable high-$\ledd$ solution, that emerges in our magnetic S-curves, is therefore a new family of solutions for a disk-corona system: radiatively efficient, radiation-pressure supported, and stabilized by magnetic pressure rather by advection.

In this Letter we show that magnetically supported disk--corona model reproduces simultaneously the Eddington ratio at which CLAGN transitions are observed to occur, and the timescale over which they are observed to unfold. Both diagnostics independently localize the instability in the inner $\sim 10$--$30\,R_{\rm Schw}$ of the flow. Sec.~\ref{sec:method} describes the model and the two diagnostics that can come out of it: timescales and S-curves. Sec.~\ref{sec:observ} confronts the model with a sample of well-characterized sources. 
Sec.~\ref{sec:discussion} discusses our results and Sec.~\ref{sec:conclusions} summarizes our conclusions. Finally, the two appendices are dedicated for describing model assumptions and limitations (App. \ref{app:limitations}), and displaying the S-curves for the other sources (App. \ref{app:more_sources}).

\section{Magnetic disk-corona model}
\label{sec:method}

We use the \texttt{diskvert} \footnote[1]{\url{https://github.com/gronki/diskvert}} code created by \citet{Gronkiewicz20,Gronkiewicz23}, that solves  a closed set of 1D vertical equations -- the hydrostatic equilibrium, energy balance, radiative transfer based on mean intensity \citep{Rozanska15}, and the magnetic-flux equation 
\citep{Begelman15}. Thanks to the relaxation method used in the algorithm, the equations are solved in a vertical domain, in the $z$ direction, in which the grid extends from the disk midplane all the way to the optically thin hot surface, through an optically thick warm corona. Details of the model, its assumptions and implementations, can be found in \citet{Gronkiewicz20, Gronkiewicz23}.

The total pressure is resolved into gas $P_{\rm gas}$, radiation $P_{\rm rad}$, and magnetic $P_{\rm mag}$ components, with a magnetic viscosity $\alpha_{\mathrm B} \equiv 2A \ P_{\rm mag}/P_{\rm tot}$, where $A$ is  the constant vertically averaged ratio of magnetic torque over magnetic pressure \citep{Salvesen16, Jiang14}. 

Heating is provided by magnetic dissipation either through the upward motion of magnetic ropes or magnetic reconnection due to the MRI dynamo. The warm corona is generated self-consistently on top of the colder disk as the result of this magnetic heating \citep{Gronkiewicz20,Gronkiewicz23}.

Each model solution derived in this work is steady and axisymmetric at fixed black hole mass $M_{\rm BH}$ in solar mass units $\msun$, accretion disk radius, $R$ in the units of $R_{\rm Schw} = 2G\mbh/c^{2}$, Eddington ratio $\ledd$ and magnetic viscosity $\alpha_{\rm B}$. In theory, \texttt{diskvert} prescribes accretion rate, $\dot{m} = \dot{M}/\dot{M}_\mathrm{Edd}$, as an input parameter. But, since they adopt constant accretion efficiency $\eta=0.083$, $\dot{m}$ and $\ledd$ can be used interchangeably. The Eddington luminosity 
has typical definition 
$L_{\rm Edd} = {4\pi GM_{\rm BH}m_{\rm p}\,c}/{\sigma_{\rm T}} $, with $G$ being the gravitational constant, $m_{\rm p}$ the proton mass, $c$ the speed of light, and $\sigma_{\rm T}$ the Thomson cross section.
A closer look on the model assumptions and limitations is presented in the App.~\ref{app:limitations}.

\subsection{Timescales}

Although the model results in a steady-state solution, the converged vertical structure derives both, the local energy content $E_t(z)$, and the local heating rate $\mathcal{H}(z)$, and their ratio defines a local thermal timescale at every height. Taking the local energy content as the sum of the enthalpies associated with the gas, radiation, and magnetic pressure components, we obtain thermal timescale

\begin{equation}
\label{eq:tth_local}
\begin{aligned}
    t_{\rm th}(z) &= \frac{E_t(z)}{\mathcal{H}(z)}, ~~~~~~ {\rm for} \\[1ex]
    E_t(z) &= \tfrac{5}{2}\,P_{\rm gas}(z) + 4\,P_{\rm rad}(z) + 2\,P_{\rm mag}(z),
\end{aligned}
\end{equation}
where each prefactor follows from the corresponding adiabatic index ($\gamma = 5/3,\,4/3,\,2$, respectively), and the steady-state condition makes heating equivalent net radiative cooling $\mathcal{H}=\Lambda$.

\begin{figure}
\centering
\epsscale{1.2}
\plotone{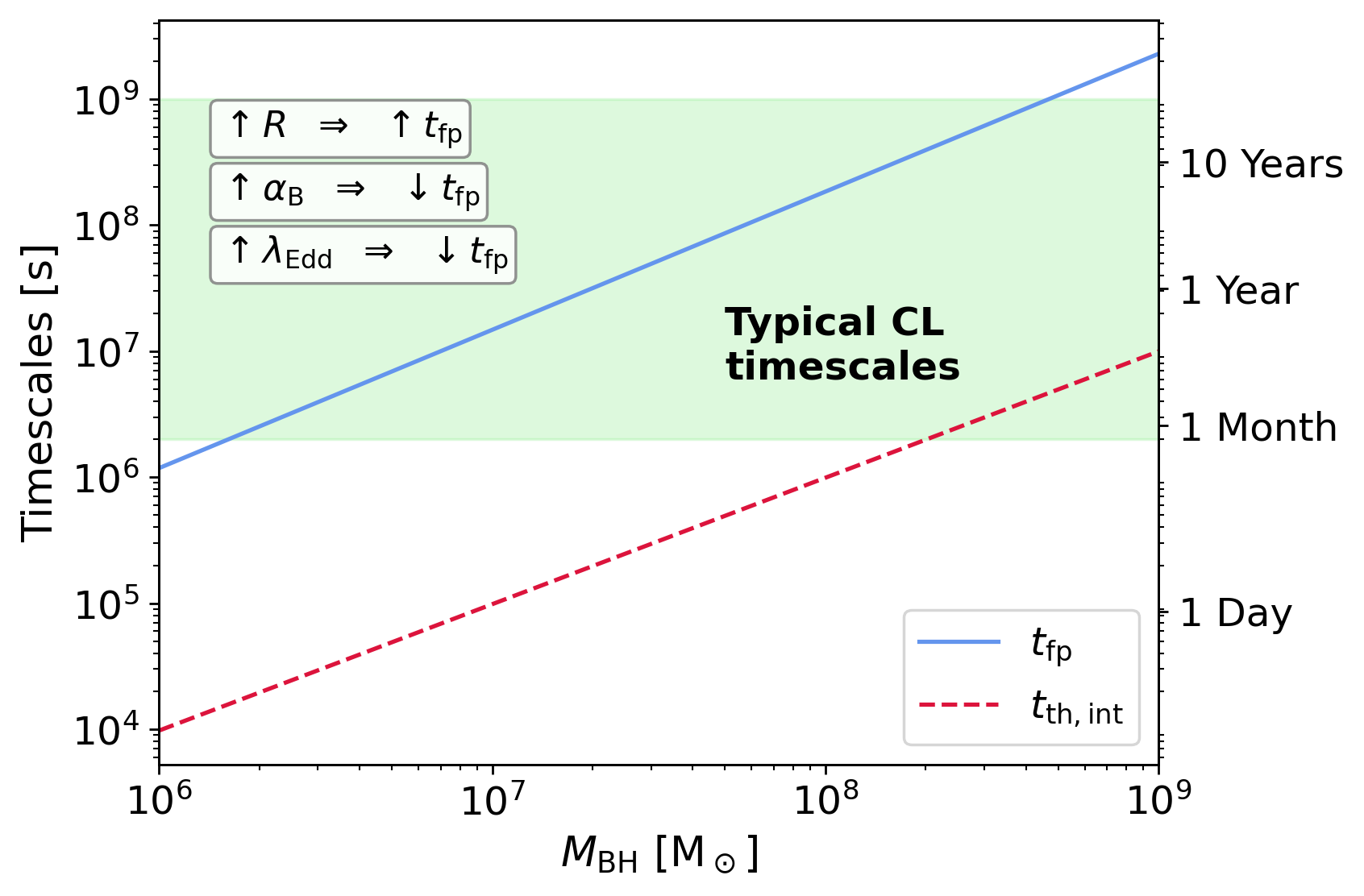}
    \caption{Integrated thermal timescale $t_\mathrm{th,int}$ (red dashed) and front propagation timescale $t_\mathrm{fp}$ (blue solid) as a function of black hole mass, computed for $\ledd = 0.01$, $R = 20\,\rschw$ and $\ab = 0.5$. The shaded band marks typical observed CLAGN transition durations. The annotation in the upper left indicates the qualitative dependence of $t_\mathrm{fp}$ on the model parameters held fixed in this figure.}
\label{fig:timescales_mass}
\end{figure}
The vertically integrated thermal timescale is then:

\begin{equation}
\label{eq:tthint}
t_{\rm th,int} \;=\; \frac{\int_{-H}^{H} E_t(z)\,dz}
    {\int_{-H}^{H} \mathcal{H}(z)\,dz},
\end{equation}
with $H$ being half of the disk scale height. As shown by the lower curve in Fig.~\ref{fig:timescales_mass}, $t_{\rm th,int}$ scales linearly with $M_{\rm BH}$, but lies within the days-to-weeks range -- too rapid to account for the duration time of CL transition.

The macroscopic state transition is set instead by the speed at which the propagation front sweeps the unstable region. Following \citet{Hameury09} the time to traverse a region of radial size $\sim R$ is the local thermal timescale geometrically amplified by the disk aspect ratio:

\vspace{-0.4em}

\begin{equation}
\label{eq:tcf}
t_{\rm fp} \;=\; t_{\rm th,int}\ \times\bigl(R/H\bigr),
\end{equation}
with $H/R$ taken self-consistently from the converged vertical solution. The upper curve in Fig.~\ref{fig:timescales_mass} shows that $t_{\rm fp}$ falls squarely in the months-to-years window for $M_{\rm BH} = 10^{6}$--$10^{9}\, \msun$, matching the observed CL-event duration for AGN discussed in Sec.~\ref{sec:observ}.

\subsection{Magnetic S-curves} 
\label{sec:scurves}

\begin{figure}[H]
\centering
\epsscale{1.15}
\plotone{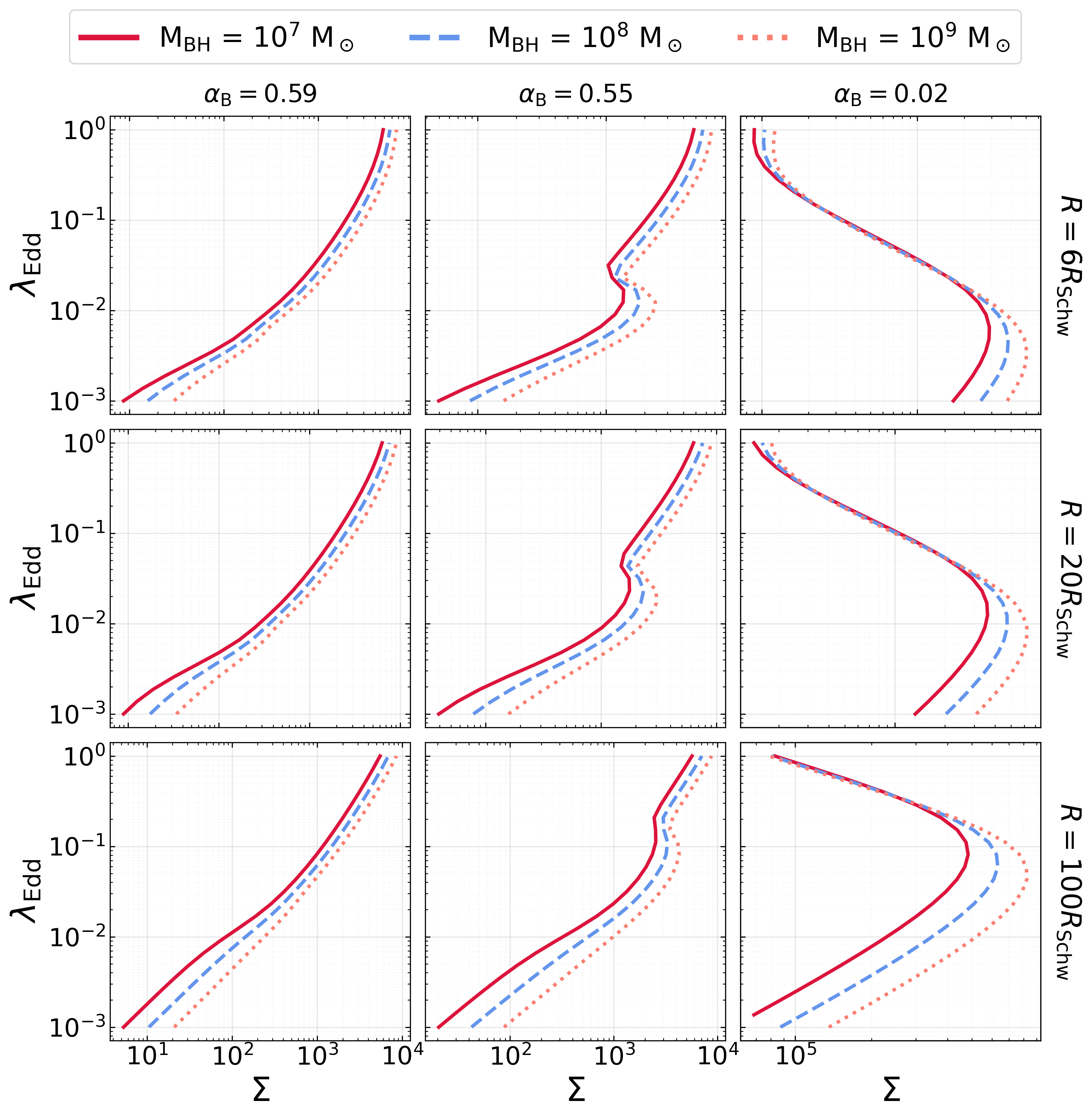}
\caption{Parametric grid of S-curves in the $\lambda_{\rm Edd} - \Sigma$ plane. Rows correspond to $R=6,\,20,\,100\,R_{\rm Schw}$ (top to bottom). Columns span the magnetic viscosity from $\alpha_{\rm B}=0.59$ (left) to $\alpha_{\rm B}=0.02$ (right). Within each panel, line colors and styles denote $M_{\rm BH} \in \{10^{7},10^{8},10^{9}\}\,\msun$, 
as marked in the box above.}
\label{fig:scurves_grid}
\end{figure}

For each set of global disk parameters $M_{\rm BH},R,\alpha_{\rm B}$, we vary $\lambda_{\rm Edd}$ and compute the converged, i.e. disk plus warm corona, column density $\Sigma$, mapping the family of steady solutions on the $\lambda_{\rm Edd}$-$\Sigma$ plane. Fig.~\ref{fig:scurves_grid} presents the resulting parametric grid. Three trends emerge.


\emph{(i) Magnetic stabilization}: at $\alpha_{\rm B} = 0.02$ the curves are highly distorted with no clear knee, and the radiation-pressure-driven instability spans a wide range of $\lambda_{\rm Edd}$; as $\alpha_{\rm B}$ increases a clean thermal-viscous knee emerges; at the maximum tested magnetic viscosity, $\alpha_{\rm B}=0.59$, the unstable middle branch disappears entirely. 

\emph{(ii) Radial dependence}: at $R = 6\,R_{\rm Schw}$ the unstable branch is deep and the S-shape is pronounced, while at $R = 100\,R_{\rm Schw}$ gas pressure dominates the vertical budget and the curves flatten - the outer disk is intrinsically more stable. 

\emph{(iii) Mass independence of the knee Eddington ratio}: varying $M_{\rm BH}$ shifts the curves laterally along $\Sigma$-axis  but leaves the vertical position of the knee $\lambda_{\rm Edd}^{\rm crit}$ essentially invariant, so $\lambda_{\rm Edd}^{\rm crit}$ is a property of the accretion physics, not of the global SMBH scale.

\begin{figure*}[!htbp]
\centering
\epsscale{1.17}
\plotone{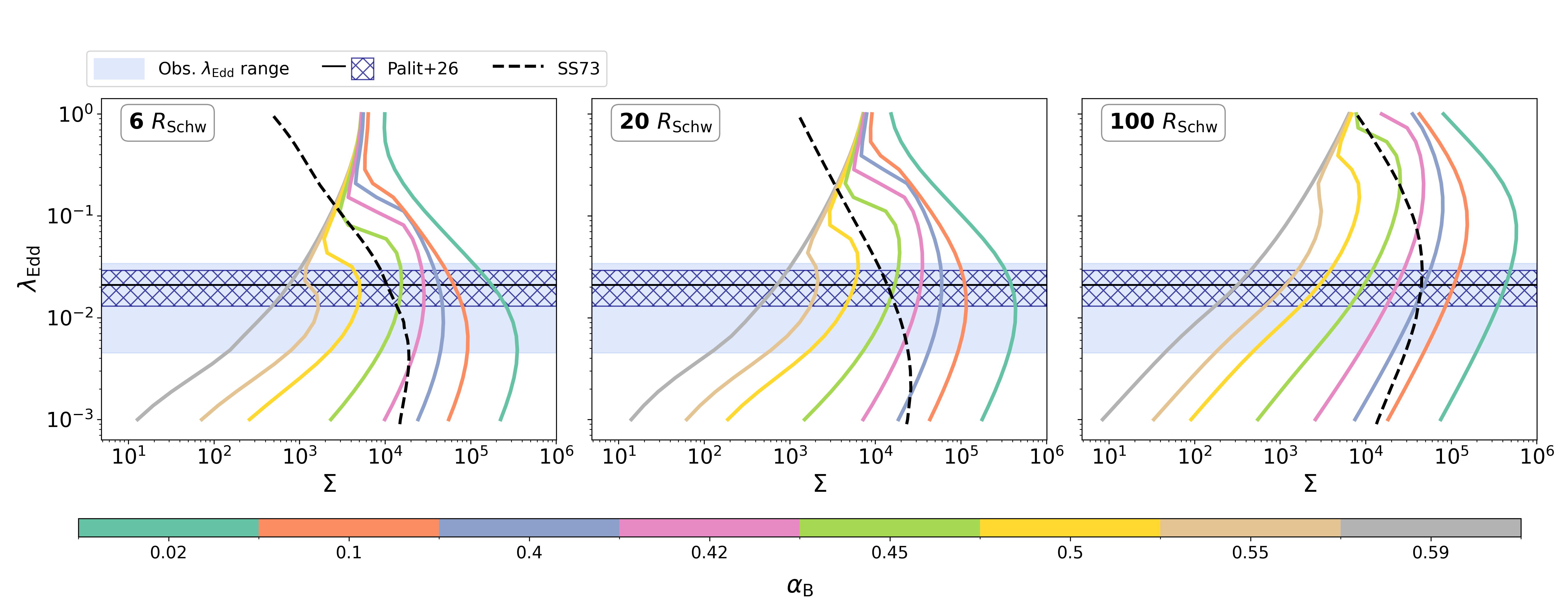}
\caption{Theoretical S-curves at the $\ledd-\Sigma$ plane, at the literature black hole mass of Mkn 590 and at $R = 6,\,20,\,100\,\rschw$, color-coded by $\ab$. Horizontal lines mark the empirical CL transition Eddington ratios from \citet{Jana25} by shaded band, and \citet{Palit26} by band with crossed, diagonal lines. The empirical thresholds intersect the theoretical knees only for $\ab \approx 0.40\text{--}0.55$ at the inner radii. Black dashed lines correspond to S-curves predicted from standard accretion theory prescribed with $\alpha = 0.1$.}
\label{fig:scurves_mkn590}
\end{figure*}

\section{Theory to Observations} 
\label{sec:observ}

To test the validity of our strongly magnetized disk models, we compare our theoretical instability thresholds directly against empirical data from a selected sample of well-documented CLAGN: Mkn 590, NGC 1566, IRAS 23226$-$3843, Mkn 1018, NGC 2617. Measured black hole masses of those sources come from reverberation mapping or single-epoch virial estimates \citep{Peterson2004,Smajic2015,Koss2022,McElroy2016,Feng2021,Wevers2019}. Empirical transition Eddington ratios $\lambda_{\rm Edd}^{\rm break}$ are taken from \citet{Jana26} for the bulk of the sample and from the dedicated study of Mkn 590 from \citet{Palit26}. The observed  duration $t_{\rm obs}$ of the CL transition for each source, together with their critical accretion rates and black hole masses are compiled from the timing literature and cited in Tab.~\ref{tab:sample}.

\subsection{S-curve test: $\lambda_{\rm Edd}^{\rm crit}$}
 
For every source, with \texttt{diskvert} code, we computed the family of S-curves at the literature $\mbh$, for $R = 6$, $20$ and $100\,\rschw$, sweeping $\ab$ from $0.02$ to $0.59$. Fig.~\ref{fig:scurves_mkn590} shows the result for Mkn 590; the analogous panels for the rest of the CLAGN sample are presented in App.~\ref{app:more_sources}.

A note on the empirical comparison is in order. The transition Eddington ratio of a CLAGN is not a single, observer-independent quantity. The hot corona photon index $\Gamma$--$\lambda_{\rm Edd}$ relation \citep[e.g.,][] {Sobolewska11, She18, Jana26} traces the photon index of the hard X-ray power law and is therefore a diagnostic of the \emph{hot} corona alone; the $\alpha_{\rm OX}$--$\lambda_{\rm Edd}$, relation where $\alpha_{\rm OX}$ index is an X-ray loudness parameter, \citep[e.g.,][]{Ruan19, Palit26} traces the optical-to-X-ray slope and is sensitive to the \emph{coupled} disk--corona system. The two diagnostics show breaks at slightly different $\lambda_{\rm Edd}$ because they probe different components of the inner flow. Our model resolves the disk and the warm corona self-consistently, but does not include the hot corona; the appropriate empirical handle for direct comparison is therefore the disk--corona one. In the absence of a $\alpha_{\rm OX}$-independent measurement for every source in our sample, we adopt the canonical CLAGN transition band $\ledd\approx 10^{-2}\text{--}2\!\times\!10^{-2}$, consistent with multiple independent analyses across the literature \citep{Stern18,Ross18,Noda18,Ruan19,Hagen24,Palit26}; for Mkn590 we also overplot the source-specific $\alpha_{\rm OX}$ measurement of \citet{Palit26}, $\log\ledd^{\rm crit} = -1.68^{+0.14}_{-0.21}$, which is the tightest single constraint currently available.

The horizontal, shaded band at Fig.~\ref{fig:scurves_mkn590} intersects the theoretical S-curves precisely at the knee of the strongly magnetized solutions, $\ab \approx 0.5 \text{--}0.55$ at $R \approx 6\text{--}20\,\rschw$. Standard, weakly magnetized models ($\ab \le 0.4$) place the knee at $\ledd \gtrsim 0.1$, an order of magnitude above the observed band: in those models, CL transitions should not occur at the luminosities at which they are actually seen. Mkn 590 sharpens this further -- the small uncertainty on the \citet{Palit26} value discriminates between magnetic viscosity configurations, and only $\ab \approx 0.55$ at $R \approx 6\text{--}20\,\rschw$ falls within the quoted $1\sigma$ band marked with crossed diagonal lines.

The point we wish to emphasize is not the precise value of the break - whether $\ledd^{\rm crit} = 10^{-2.0}$ or $10^{-2.47}$ depends on the diagnostic, the bolometric correction, and the sample. The remarkable observational fact is that a break \emph{exists} in this narrow band, and is recovered by independent analyses using independent indicators across heterogeneous CLAGN samples. Our model offers a physical explanation: this is the Eddington ratio at which the upper unstable radiation pressure-dominated branch of the magnetized disk's S-curve disappears, forcing the inner flow off its stable disk solution. This set of solutions corresponds to obtaining warm Comptonization and a luminous warm corona \citep{Gronkiewicz23}. The location of the break is set determined by the value of $\ledd$ for which magnetic support can no longer prevent the radiation-pressure-driven instability -- a value that depends on the magnetic viscosity $\ab$ and on the radius within the inner disk, but not on the absolute scale of the SMBH (cf. the mass-independence of the knee in Fig.~\ref{fig:scurves_grid}). Below that break, the disk enters once again a stable branch, but now it is gas-pressure supported and the warm corona is suppressed.


\subsection{Timescale test: $t_{\rm fp}$}
 
\begin{figure}[!htbp]
\centering
\epsscale{1.17}
\plotone{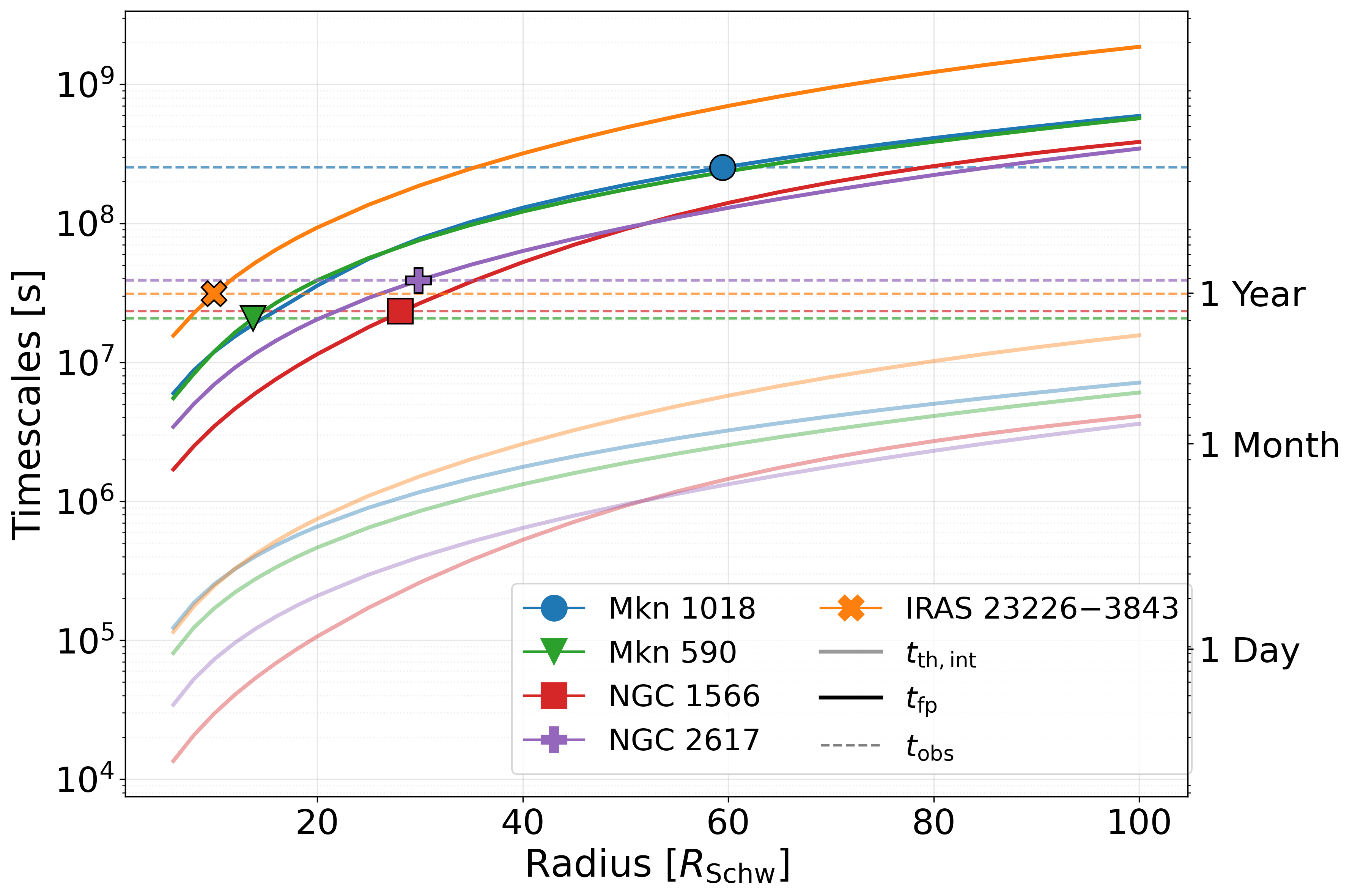}
\caption{Vertically integrated thermal timescales $t_{\rm th,int}$ (faded lower curves) and front propagation timescales $t_{\rm fp}$ (solid upper curves), as a function of disk radius for the five CLAGN in our sample given by different colours and markers. Horizontal dashed lines mark the empirically observed transition duration, and the solid dots indicate the radius at which $t_{\rm fp}$ matches $t_{\rm obs}$.}
\label{fig:tscale_sources}
\end{figure}

We then asked whether the same models reproduce the observed transition durations. For each source, from Eqs.~\ref{eq:tthint} and \ref{eq:tcf}, we compute integrated timescales $t_{\rm th,int}$ and $t_{\rm fp}$, as a function of $R$ at the source-specific empirical transition Eddington ratio of \citet{Jana25} and at the source-specific best-fit $\ab$ inferred from the S-curve comparison (Tab.~\ref{tab:sample} sixth column). Only for the case of Mkn 590 we used the value of $\ledd^{\rm break}$ constrained by \citet{Palit26}. Fig.~\ref{fig:tscale_sources} demonstrates that modelled timescales, derived at each distance from SMBH from source-dependent vertical structure calculations of magnetized disk, plotted by solid lines, intersect the observed $t_{\rm obs}$ overplotted as horizontal dashed lines.

It is clear that the integrated thermal timescale is universally an order of magnitude or more too short, while the front propagation timescale crosses the observed durations at $R \approx 10\text{--}60\,\rschw$, depending on the source. The intersection radius $R_{\rm tr}$, marked by solid dots at the plot, and derived from the timescale's match (Tab.~\ref{tab:sample}, last column) is consistent, source by source, with the inner range $R \approx 6\text{--}20\,\rschw$ favoured by the S-curve fit -- bearing in mind that $R_{\rm tr}$ from $t_{\rm fp}$ is the radius across which the front \emph{propagates}, while the S-curve fit identifies the radius at which the instability is \emph{ignited}. Both diagnostics therefore agree that CL events are dynamically restricted to the inner accretion flow.
 
\section{Discussion} \label{sec:discussion}

In this Letter, we demonstrate, that two independent diagnostics -- the S-curve knee in $\ledd$, and the front propagation timescale in seconds -- converge on the same physical picture. CL transitions are triggered when the Eddington ratio of the inner disk, where the disk is sustained against radiation-pressure collapse by a strong magnetic component ($\ab \approx 0.40\text{--}0.55$), drops below an Eddington ratio of the order of $10^{-2}$, falls off the upper branch of its S-curve, and propagates the resulting front through the inner $\sim 20\text{--}60\,\rschw$ on a $R/H$-amplified thermal timescale. The same model that drops the knee to the observed $\ledd^{\rm crit}$ predicts a transition duration of months to years, because both effects share the same physical lever: vertical magnetic support.


\begin{figure}[!htbp]
\centering
\epsscale{1.17}
\plotone{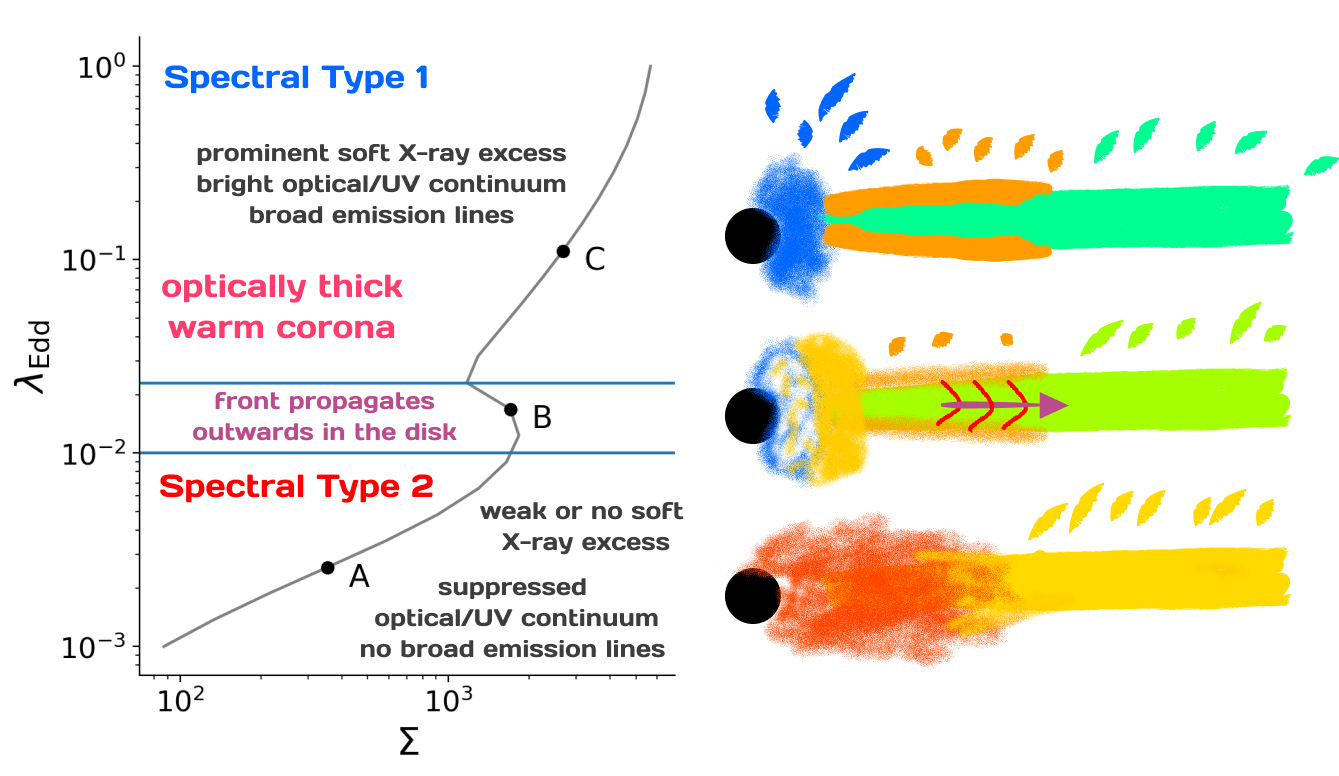}
\caption{The accretion states of a CLAGN across the magnetic S-curve. The left panel shows the thermally stable and unstable branches for a highly magnetized disk in the $\ledd - \Sigma$ plane. The corresponding cross-sectional figures detail the disk and warm corona evolution. A stable Type-1 with a strong soft X-ray excess (C) destabilizes at the critical knee (B), where a front propagates outward on the timescale $t_{\rm fp}$. The transition concludes on the radiatively inefficient lower branch (A), characterized by a vanished warm corona and a dim Type-2 spectrum.}
\label{fig:drawing}
\end{figure}

A natural by-product of this picture is a re-reading of the Type-1/Type-2 dichotomy \citep{Antonucci93} and its variability. The radiation pressure unstable branch in thermal-viscous S-curve of a standard thin disks, becomes stable when magnetic pressure is large enough (Fig.~\ref{fig:scurves_grid}), thus enabling for a new limit-cycle to arise. Depending on the value of Eddington ratio, the magnetized disk may oscillate
between: the upper stable branch of the S-curve that corresponds to a radiatively efficient, geometrically thin solution with a luminous warm corona at the photosphere -- a configuration that reproduces the soft X-ray excess, and the broad lines from BLR characteristic of bright Type-1 spectra; and the lower branch where the inner flow is dominated by a hot corona with optical/UV continuum suppressed due to disk truncation at larger radii, with no warm corona and with narrow lines from BLR - all being characteristic to Type-2 spectra. The position of both accretion states is reflected at magnetic S-curve and shown in  Fig.~\ref{fig:drawing}. The intermediate, unstable branch depicts the front propagation outward on the timescale $t_{\rm fp}$.


The novelty of our approach lies in proposing a magnetized disk and showing that the same vertical-structure code that delivers the front propagation timescale also delivers the global S-curve, so the trigger condition $\ledd^{\rm crit}$ and the propagation time $t_{\rm fp}$ are derived from a single self-consistent set of solutions rather than parametrized independently.

\section{Conclusions}  \label{sec:conclusions}

We have presented that a vertically resolved, magnetically supported disk-corona model can simultaneously satisfy two empirical constraints of Changing-Look AGN which standard $\alpha$-disk theory cannot meet:

\begin{enumerate}
\item Strongly magnetized configurations, $\ab \approx 0.40$--$0.55$, push the thermal-viscous S-curve knee from down to $\ledd \approx 0.01\text{--}0.03$, matching the empirical CL transition Eddington ratios of Mkn 590, NGC 1566, IRAS 23226$-$3843, Mkn 1018 and NGC 2617. In the same configurations, the front propagation timescale $t_{\rm fp} $ falls in the months-to-years range for SMBH masses, in agreement with the observed transition durations of the sample.
\item The S-curve diagnostic and the timescale diagnostic both localize the instability inside $\sim 10\text{--}30\,\rschw$, i.e.\ in the inner accretion flow for the considered black hole masses.
\item The magnetic S-curve introduces a new stable upper branch that corresponds to high-luminosity solutions, which is separated by an unstable middle segment from the lower stable branch. This offers a new limit cycle in which the disk oscillates between the two stable branches. This cycle is driven by the radiation-pressure instability of the magnetised inner disk and is operating at the empirically observed Eddington ratios. A time-dependent calculation that follows the cycle explicitly is the natural next step and will be presented in a forthcoming work.
\item The knee Eddington ratio is essentially mass-independent, so CL events are predicted to occur at a universal $\ledd^{\rm crit}$ across the entire SMBH mass range probed.
\item Mkn 590 provides the most stringent single test: the tight $\ledd^{\rm break}$ from \citet{Palit26} is reproduced only by $\ab \approx 0.55$ at $R \approx 6$--$20\,\rschw$.
\end{enumerate}

\begin{acknowledgments}
\end{acknowledgments}
The research leading to these results have received funding by the EU HORIZON-MSCA-2023-DN Project 101168906  “TALES: Time-domain Analysis to study the Life-cycle and Evolution of Supermassive black holes”. MK acknowledges full and AR partial funding. DL acknowledges the Czech Science Foundation (GA\v{C}R) project No. 25-16928O.




\bibliography{clagn}{}
\bibliographystyle{aasjournalv7}

\appendix

\renewcommand{\thetable}{B\arabic{table}}
\setcounter{table}{0}
\vspace{-1.0cm}
\begin{deluxetable}{lcccccc}[h]

\tablecaption{Sample of CLAGN used in this work with their redshift given in column 2. Black hole mass, the observed breakpoint Eddington ratio $\lambda_{\rm Edd}^{\rm break}$, from $\Gamma$ or $\alpha_{\rm OX}-\ledd$ relation and the measured time duration of CL event, are given in columns 3, 4, and 5, respectively. Columns 6 and 7 show the magnetic viscosity and the radius at which our $t_\mathrm{fp}$ matches the $t_\mathrm{obs}$. In the case of Mkn 1018 there is not enough data coverage to conclude a breakpoint value.}
\tablehead{
\colhead{Source} & \colhead{$z$} & \colhead{$M_{\rm BH} / 10^{7}\,\msun$} & \colhead{$\log \lambda_{\rm Edd}^{\rm break}$} & \colhead{$t_{\rm obs}$} & \colhead{optimal fit $\ab$} & \colhead{$R_{\rm tr}/\rschw$}
}
\startdata
Mkn 590           & 0.0260 & $4.75\pm0.74$\,$^{a}$ & $-1.68^{+0.14}_{-0.21}$\,$^{f}$ & $\sim 8$\,mo $^{f}$  & 0.55 & 14 \\
NGC 1566          & 0.0047 & $0.53\pm0.29$\,$^{b}$ & $-2.58 \pm 0.09$\,$^{g}$                  & $\sim 9$\,mo $^{h}$& 0.5 & 28 \\
IRAS 23226$-$3843 & 0.0350 & $6.76^{+1.37}_{-1.14}$\,$^{c}$ & $-2.25 \pm 0.16$\,$^{g}$         & $\sim 1$\,yr $^{i}$& 0.55 & 10 \\
Mkn 1018          & 0.0420 & $7.94$\,$^{d}$         & $-$                  & $\sim 8$\,yrs  $^{j}$& 0.5 & 59 \\
NGC 2617          & 0.0142 & $2.1\pm0.4$\,$^{e}$    & $-2.48 \pm 0.06$\,$^{g}$                  & $\sim 15$\,mo $^{k}$& 0.5 & 30 \\
\enddata
\tablecomments{References: $^{a}$\citet{Peterson2004}; $^{b}$\citet{Smajic2015}; $^{c}$\citet{Koss2022}; $^{d}$\citet{McElroy2016}; $^{e}$\citet{Feng2021}; $^{f}$\citet{Palit26}; $^{g}$\citet{Jana26}; $^{h}$\citet{Parker19}; $^{i}$\citet{Kollatschny26}; $^{j}$\citet{Noda18}; $^{k}$\citet{Oknyansky17}.}

\label{tab:sample}
\end{deluxetable}

\vspace{-1.7cm}

\section{Model assumptions and limitations}
\label{app:limitations}


The S-curves presented in this Letter are evaluated at radii $R \geq 6\,\rschw$, i.e.\ outside or at the innermost stable circular orbit (ISCO) of a Schwarzschild black hole. This is not a numerical limitation but a physical feature of the local heating profile adopted by \texttt{diskvert}. The code implements the standard \citet{SS73} dissipation per unit area,

\begin{equation}
\label{eq:NT_profile}
F_{\rm acc}(R) \;=\; \frac{3 G M \dot M}{8\pi R^{3}}\,
\Bigl[\,1 - \sqrt{R_{\rm ISCO}/R}\,\Bigr],
\end{equation}
with $R_{\rm ISCO} = 3\,\rschw$ fixed (Schwarzschild, $a = 0$). The bracketed factor enforces the no-torque inner boundary condition: matter free-falls inside the marginally stable orbit and the local heating rate vanishes at $R = R_{\rm ISCO}$.

A direct consequence is that $F_{\rm acc}(R)$ is non-monotonic. Setting $dF_{\rm acc}/dR = 0$ gives a maximum at $R = (49/36)\,R_{\rm ISCO} \approx 4\,\rschw$. Beyond this peak, $F_{\rm acc}$ falls in either direction -- as $R^{-3}$ outward, and as $(1 - \sqrt{R_{\rm ISCO}/R})$ inward. Because the S-curve knee location depends on the local heating rate, the knee Eddington ratio $\ledd^{\rm crit}$ inherits this non-monotonicity: two S-curves evaluated at radii sitting on opposite sides of the peak (for example $R \approx 4\,\rschw$ and $R \approx 10\, \rschw$) yield essentially the same $\ledd^{\rm crit}$. Evaluating S-curves inside $R = 6\,\rschw$ therefore does not push the knee to lower Eddington ratios than those already obtained near the dissipation peak. We adopt $R = 6\,\rschw$ as the inner edge of the grid because it sits comfortably outside the ISCO and close enough to the maximum that the attainable minimum of $\ledd^{\rm crit}$ is essentially recovered.

A non-zero spin would lower $R_{\rm ISCO}$ and shift the peak inward in absolute units, but the same non-monotonic structure of $\ledd^{\rm crit}(R)$ would persist. The current implementation hardcodes $a = 0$, and SMBH rotation is not considered.

The model is steady-state and one-dimensional; the front is treated through the geometric $R/H$ amplification rather than by an explicit time-dependent calculation. Magnetic viscosity is held uniform across the disk; in reality the viscosity may itself depend on $R$ \citep{Abramowicz26} and on the accretion history. The required magnetic viscosity, $\ab \approx 0.5$, is consistent with the values reported in simulations of magnetically stabilized disks \citep{Lancova19,Mishra20,Begelman24}.


The code \texttt{diskvert} resolves the cold disk and the warm corona self-consistently, but does not include a hot ($kT \sim 100$ keV), optically thin corona of the type required to produce the hard X-ray power law observed in AGN. The warm-corona model addresses the soft X-ray excess and the disk-corona thermodynamic coupling, while the hard X-ray power law involves a distinct phase of the inner flow whose geometry and energetics remain debated \citep{Fabian15}. This choice has a direct consequence for how the model is compared to data. The empirical $\Gamma$--$\ledd$ relation \citep{Sobolewska11,She18,Jana25,Jana26} traces the photon index of the \emph{hot} corona and is therefore not the appropriate diagnostic for a model that does not include this component. The $\alpha_{\rm OX}$--$\ledd$ relation, sensitive to the coupled disk-corona system, is the correct empirical handle and is what we use for direct comparison (Sec.~\ref{sec:observ}).

\section{Source-specific S-curves} 

\label{app:more_sources}

\begin{figure*}[!htbp]
\centering
\epsscale{1.17}
\plotone{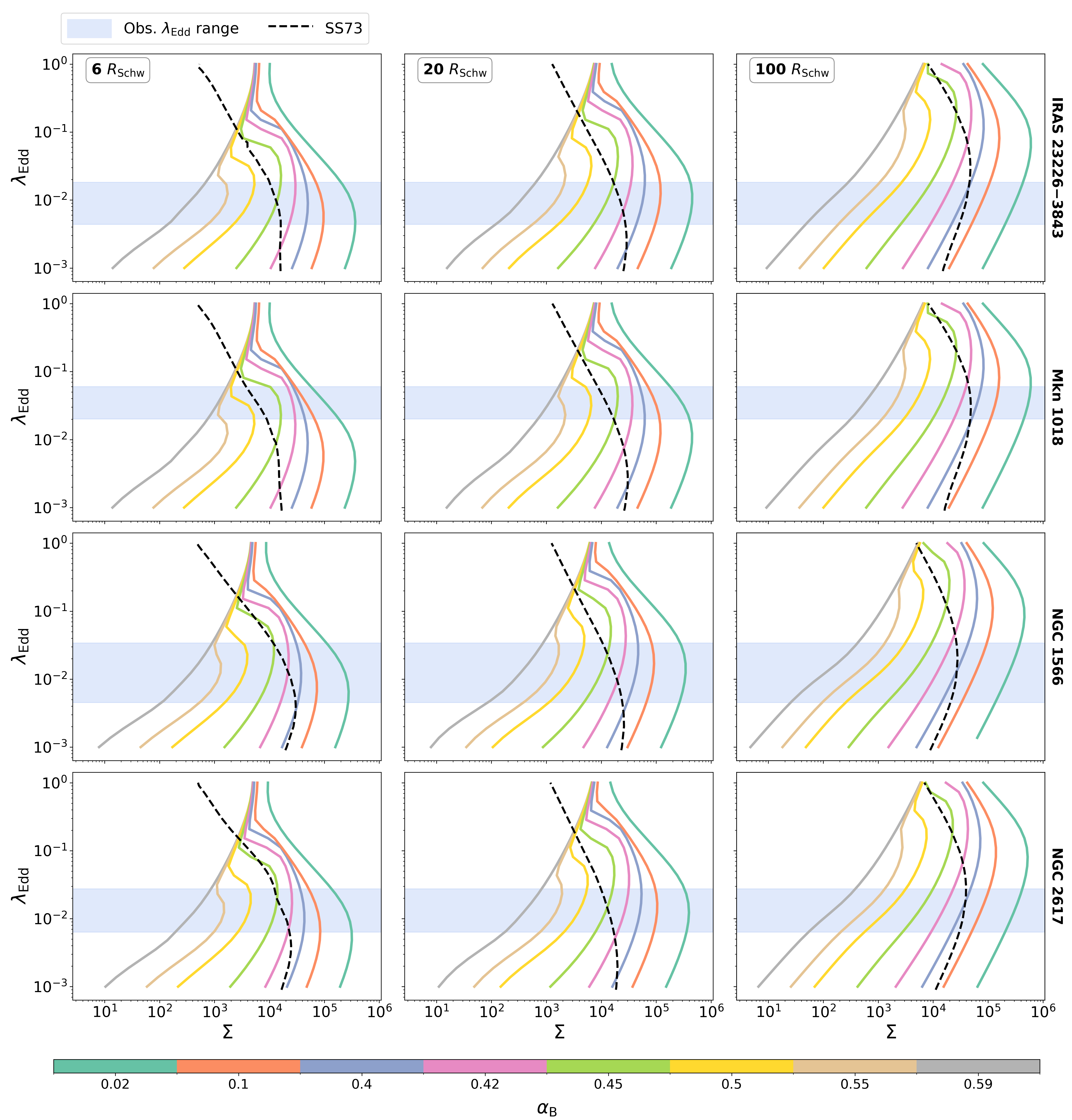}
\caption{Same as in Figure \ref{fig:scurves_mkn590}, but for NGC 1566, NGC 2617,  Mkn 1018 and IRAS 23226-3843.}
\label{fig:scurves_more_sources}
\end{figure*}

Fig.~\ref{fig:scurves_more_sources} extends the analysis of Sec.~\ref{sec:observ} to other CLAGN as: NGC 1566, NGC 2617,  Mkn 1018 and IRAS 23226-3843. The same conclusions hold: the empirical $\ledd^{\rm crit}$ is reproduced only for $\ab \approx 0.40$--$0.55$ at the inner radii.

\end{document}